\documentclass[letterpaper,twocolumn]{article}

\usepackage[T1]{fontenc}

\usepackage{geometry}
\usepackage{setspace}

\usepackage{achemso}

\usepackage{graphicx}
\usepackage{float}
\newfloat{scheme}{htbp}{los}
\floatname{scheme}{Scheme}
\floatname{chart}{Chart}
\newfloat{graph}{htbp}{loh}

\usepackage{chemformula} 
\usepackage[version = 4]{mhchem} 

\usepackage{authblk}
\author{Yuri B. Kudasov}
\affil{Sarov Physics and Technology Institute, National Research Nuclear University "MEPhI", Sarov, Russia}

\title{Unusual half-metallic state in unconventional magnets}
\date{*Email: yu\_kudasov@yahoo.com}

\begin{document}

\maketitle

\begin{abstract}
Half-metallic ferromagnets exhibit a gap in the density of states for one spin projection while remaining gapless for the opposite spin. We show that in helimagnets an unusual half-metallic state can exist, where the spin projection that experiences the gap is determined by the direction of the wave vector. This state originates from the nontrivial topology of the band structure, specifically from the dispersion forming a multi-sheeted covering over the Brillouin zone. We present two-dimensional tight-binding models for $p$-wave and $f$-wave half metals. These structures can be realized in crystalline and van der Waals systems. The complex band structure of the unusual $p$-wave half metal in nanostructures is also discussed. In a quantum well, a standing wave with a helical spin structure is formed, and a persistent spin current exists.
\end{abstract}

\section*{Keywords}

half metal, topological band structure, anisotropic gap, altermagnets, helimagnets, tight-binding model, quantum well, spin current.

\section{}

\begin{figure*}
	\; \;\; \;\; \;	\includegraphics[scale=0.38]{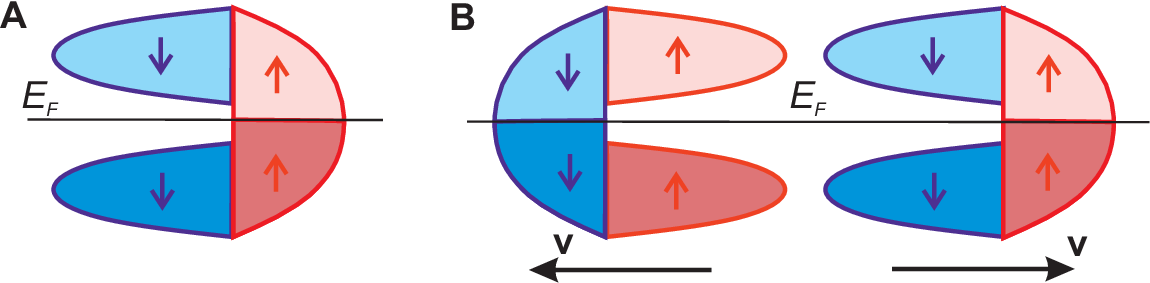}  \; \; \;
	\includegraphics[scale=0.40]{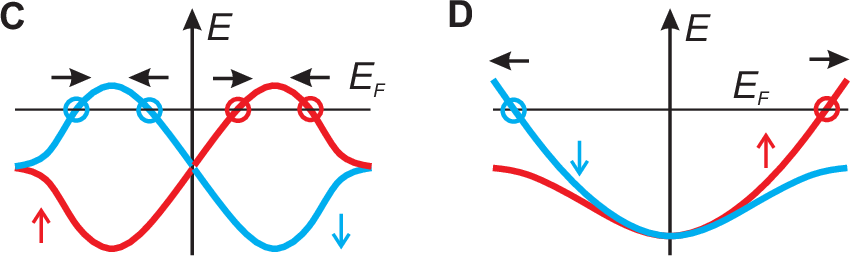} 
	\caption{\label{f1} (A) A schematic view of the DOS for the ordinary ($s$-wave) half-metallic ferromagnets. (B) The same for the 1D $p$-wave half metal. The horizontal arrows denote the left- and right-moving electrons. (C) A spin-split band. (D) A pair of topologically nontrivial bands.}
\end{figure*}

The most fundamental classification of band structures in crystalline solids is based on the position of the Fermi level relative to the electronic band gap. This criterion distinguishes metals, semiconductors, and insulators \cite{Ashcroft}. A further refinement of this framework for magnetic materials has led to the identification of half-metallic ferromagnets \cite{Katsnelson,Fong} and spin-gapless semiconductors \cite{Yue}---systems that are gapless for electrons of one spin direction while exhibiting a gap for the opposite spin (Fig.~\ref{f1}A). These materials hold significant promise for spintronics applications, as they can enable nearly full spin polarization of the injected current in nanostructures \cite{Fong,Attema}. 

The first question raised in this Letter is whether a one-dimensional band structure can exist such that, for right-moving particles, there is a gap for spin-down electrons but a gapless state for spin-up electrons, while for left-moving particles the opposite holds---a gapless state for spin-down electrons and a gap for spin-up electrons---as illustrated in Fig.~\ref{f1}B. We refer to such a structure as an unusual $p$-wave half metal. Furthermore, we ask whether this band structure can be generalized to two- and three-dimensional systems.

To induce a gap in the density of states (DOS) of the type shown in Fig.~\ref{f1}B, a band structure with wave-vector-dependent spin splitting is required. Such splitting can arise from spin-orbit coupling or in altermagnets \cite{Smejkal}. Several non-equivalent definitions of altermagnets exist \cite{Smejkal,Mazin,Cheong}. Here, we adopt the definition that describes them as zero-magnetization systems with broken parity-time symmetry. This leads to $\mathbf{k}$-dependent, nonrelativistic spin-split bands. For non-collinear magnetic structures, we also use the term "unconventional magnets" \cite{Yuan,Brekke,Ezawa}.

Since the particle velocity is related to the group velocity of the wave packet by  $\mathbf{v} = \partial \varepsilon / \partial \mathbf{k}$ \cite{Ashcroft}, a typical band spin splitting does not give rise to the $p$-wave half-metallic state, as illustrated in Fig.~\ref{f1}C. However, spin-split bands that are non-periodic within the magnetic Brillouin zone can exist in helical magnetic structures \cite{KudasovJETPL2,KudasovPRB}. The band structure of such systems has been studied for a long time using various theoretical approaches, including perturbation analysis \cite{Dzyaloshinskii}, the theory of spin-space groups \cite{Brinkman}, exact solutions \cite{Calvo}, the generalized Bloch theorem \cite{Sandratskii,KudasovJETPL2}, and others \cite{Calvo2,Kudasov,KudasovAPL}. It was found that a gapless band structure is formed \cite{Dzyaloshinskii}, which arises from nontrivial topology, as was recently proved \cite{KudasovAPL}. In contrast with topological insulators and semimetals \cite{Hasan,Armitage}, this structure is not related to the Berry phase \cite{Berry}. In this case, a group of individually non-periodic bands appears. At the same time, the group itself is periodic within the magnetic Brillouin zone as a whole. From a topological point of view, this forms a multi-sheeted covering space. The number of sheets and their sequence order (monodromy) are topological invariants \cite{KudasovAPL}. It should also be noted that the band structure of commensurate helical magnetic systems exhibits a Kramers-like symmetry and degeneracy \cite{KudasovPRB}.

To study prototypes of unusual $p$- and $f$-wave half metals, one can consider the following tight-binding model on a 2D lattice:
\begin{eqnarray}
	\hat{H} =	-\sum_{\langle i,j\rangle, \sigma}\left(t_{ij} \hat{a}^\dagger_{i\sigma} \hat{a}_{j\sigma} + \text{h.c.}\right) 
	- \mathbf{h}_{i} \sum_{i,\sigma,\sigma^\prime} \hat{a}^\dagger_{i\sigma} \hat{\boldsymbol{\tau}} \hat{a}_{i
		\sigma^\prime}  \label{Htb}
\end{eqnarray}
where $\hat{a}^\dagger_{i\sigma} (\hat{a}_{i\sigma})$ is the creation (annihilation) operator for an electron with spin projection $\sigma = \uparrow, \downarrow$ at the $i$-th site, $\hat{\boldsymbol{\tau}}$ are the Pauli matrices, and $\mathbf{h}_{i}$ is the (effective) magnetic field at the $i$-th site. The notation $\langle \ldots \rangle$ denotes the sum over nearest-neighbor pairs.

\begin{figure*}
	\;\; \;  \;	\includegraphics[scale=0.5]{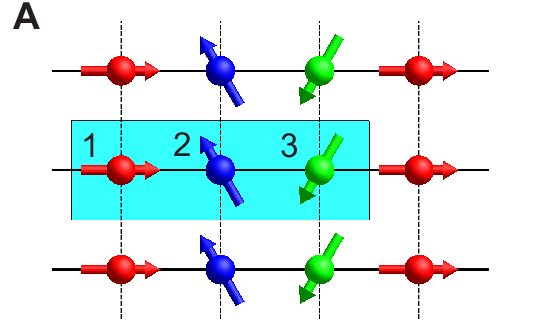}   \;
	\includegraphics[scale=0.17]{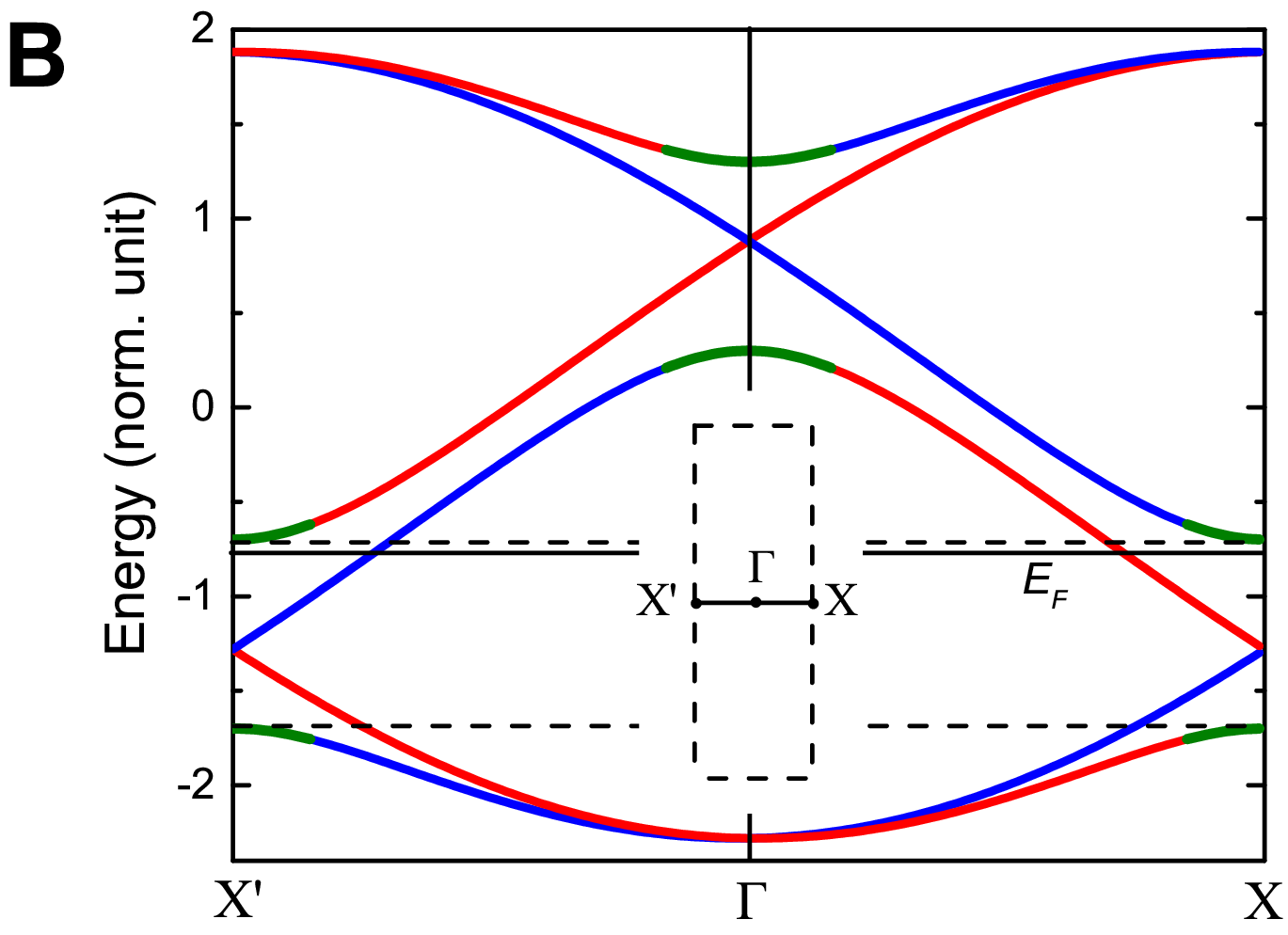}     \; \; \;
	\includegraphics[scale=0.45]{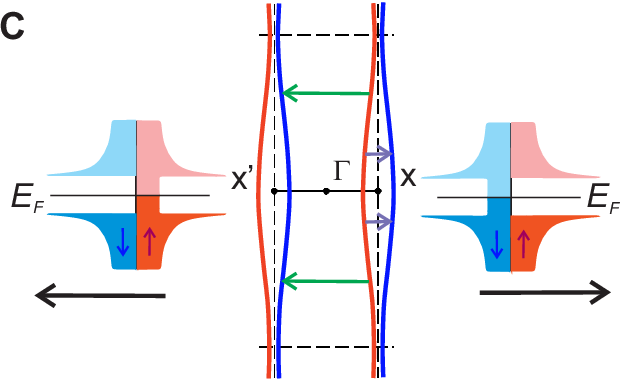} 
	\caption{\label{f2} (A) Rectangular lattice with 120$^\circ$ magnetic order. The sublattices are shown by color and numbers. The supercell is indicated by the filled rectangle. (B) Band structure along the X'-$\Gamma$-X path ($t=0.1$, $|\mathbf{h}_{i}|=h_0=0.5$). The average spin projection onto the $z$-axis is indicated by color: the red and blue for predominantly spin-up and spin-down states, respectively; the green denotes strongly mixed states, i.e., when $|\langle \hat{\tau}_z\rangle|<1/2$. (C) The Fermi surface for the Fermi level at $E_{F}=-0.75$ ($p$-wave half-metallic state). The DOS for the left- and right-moving electrons is schematically shown. The green and purple arrows show the backward and umklapp scattering vectors, respectively.}
\end{figure*}

Let us begin with a three-sublattice model featuring a 120$^\circ$ magnetic order on the rectangular lattice, as shown in Fig.~\ref{f2}A. The vectors $\mathbf{h}_{i}$ lie in the $xy$-plane, and the angle between them at different sublattices is 120$^\circ$. The nearest-neighbor hopping is assumed to be strongly anisotropic: along the $x$-axis, $t_{ij}=1$, while along the $y$-axis, $t_{ij}=t \ll 1$, as schematically indicated by solid and dashed lines in Fig.~\ref{f2}B, respectively. The dispersion curves along the X$^\prime$-$\Gamma$-X line in the magnetic Brillouin zone resemble those of the one-dimensional model with 120$^\circ$ order, up to a constant energy shift, as shown in Fig.~\ref{f2}B. The two lowest bands intersect at the $\Gamma$ point. Although the two bands are not individually periodic, the set of three bands as a whole is periodic. Finally, if the Fermi level lies within the range indicated by the dashed lines in Fig.~\ref{f2}B, the $p$-wave half-metallic state is realized.

An example of the Fermi surface in the $p$-wave half-metal regime is shown in Fig.~\ref{f2}C. The spin polarization of the sheets is indicated by color in the same manner as in Fig.~\ref{f2}B. This band structure should result in unusual transport properties. Backward scattering by non-magnetic impurities and defects (without a spin flip) is almost totally blocked because of the high spin polarization (about 98\% in Fig.~\ref{f2}C). The same can be said about the electron-phonon umklapp scattering at low temperatures. Therefore, one can expect extremely high conductivity along the $\Gamma$-X direction in such a material. 

\begin{figure*}
	\; \; \; \; \;\; \;	\includegraphics[scale=0.47]{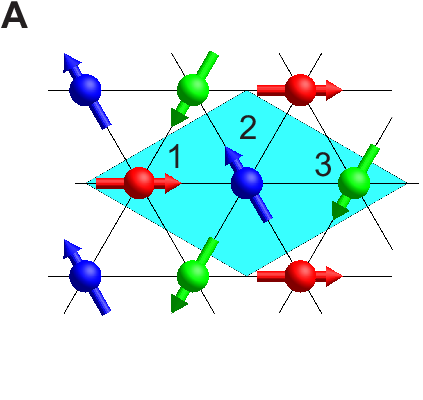}  \; \; \;
	\includegraphics[scale=0.2]{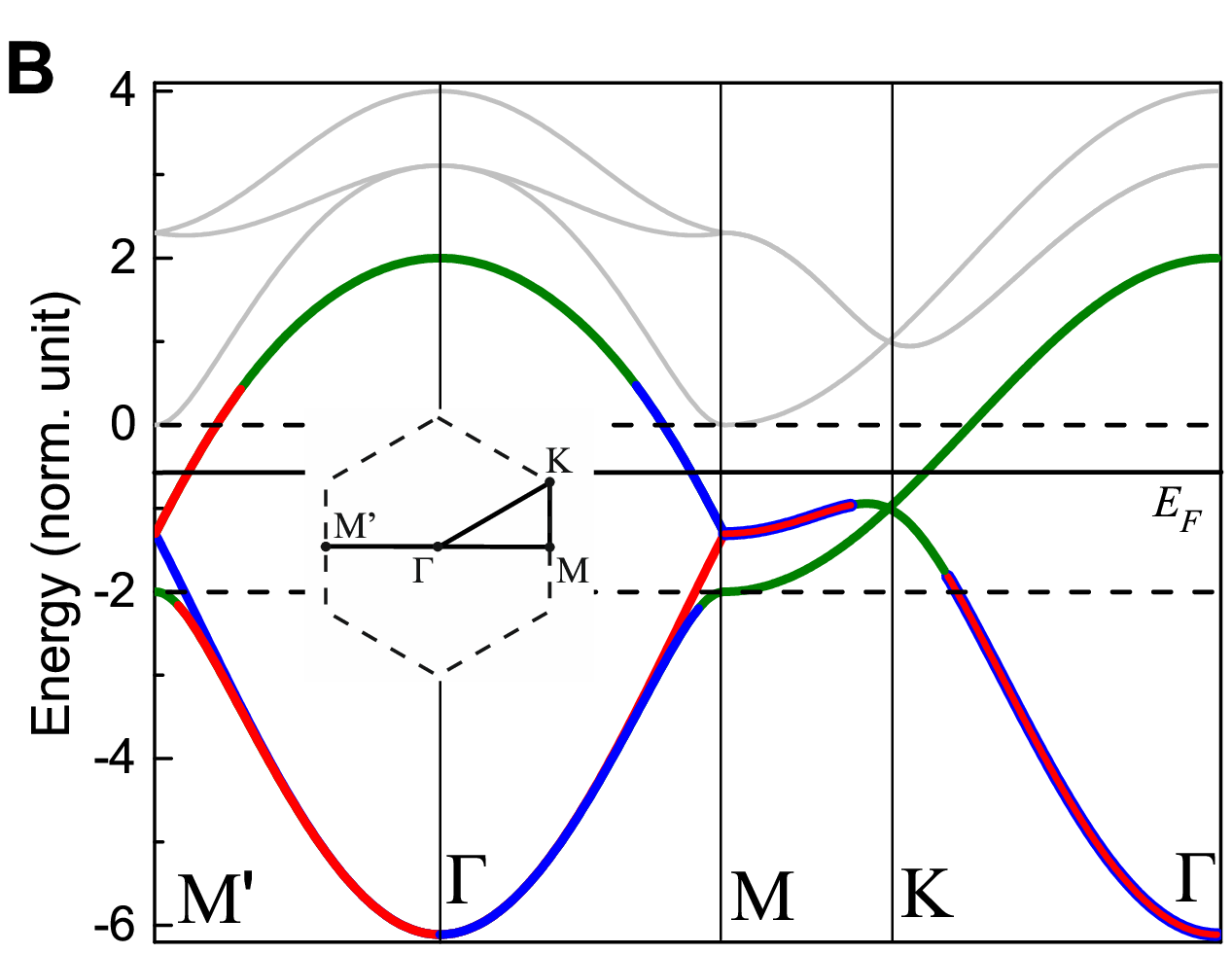}  \; \; \;
	\includegraphics[scale=0.39]{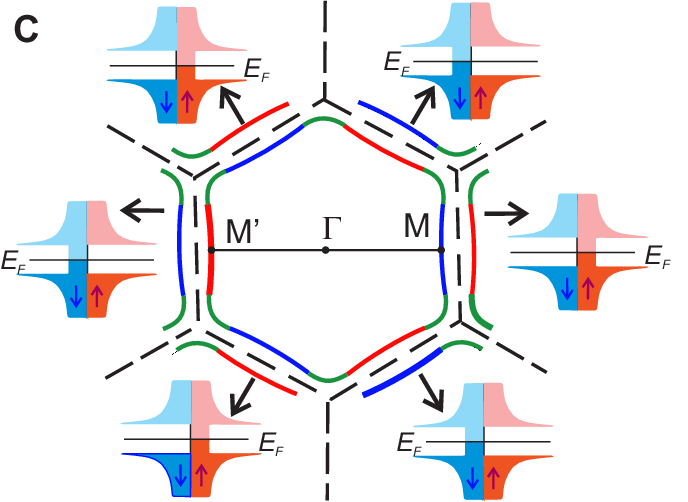} 
	\caption{\label{f3} (A) Triangular lattice with 120$^\circ$ magnetic order shown in the same manner as in Fig.~\ref{f2}A. (B) Band structure along the path shown in the inset ($h_0 = 1$). The average spin projection is indicated as in Fig.~\ref{f2}B.  (C) Fermi surface for the Fermi level at $E_{F} = -0.6$ ($f$-wave half-metallic state). The DOS for electrons moving along different directions is schematically shown.}
\end{figure*}

Another possibility for realizing the unusual 2D half metal is a hexagonal structure with 120$^\circ$ ordering, shown in Fig.~\ref{f3}A. This system is described by the same Hamiltonian (\ref{Htb}), whose explicit matrix form is presented in section I of the Supporting Information (SI). The band structure and its topology were recently discussed \cite{KudasovAPL}. The dispersion curves along the M'-$\Gamma$-M-K-$\Gamma$ path are shown in Fig.~\ref{f3}B. The M'$-\Gamma$-M segment has qualitatively the same structure of lower dispersion branches as that in Fig.~\ref{f2}B. There are band gaps for spin-down particles moving to the right along this direction and spin-up particles moving to the left. An example of the Fermi surface calculated earlier \cite{KudasovAPL}, which corresponds to the unusual half metal, is shown in Fig.~\ref{f3}C.
It should be noted that after a rotation by $\pi/6$, band gaps for spin-up and spin-down particles are interchanged, i.e., this is the $f$-wave half-metallic state. Along the $\Gamma$-K segment, the gaps for both spins collapse.

The  $f$-wave half-metallic phase should also have an anomalously high conductivity for the same reason as the $p$-wave one. Other unconventional transport properties in such a band structure, for example, nonreciprocity under an external magnetic field and unconventional anomalous Hall effect, were discussed earlier \cite{KudasovArX}.

Previously, the metallic delafossite \ch{PdCrO2} was considered a possible candidate for this type of band structure \cite{Kudasov,KudasovPRB}. This substance is a hexagonal layered material with 120$^\circ$ ordering of chromium ions in the layers \cite{Mackenzie}. Conductivity is provided by palladium layers alternating with \ch{CrO2} interlayers. Therefore, a 120$^\circ$ effective molecular field should be induced in the palladium layers. The Fermi surface obtained by model calculations \cite{KudasovPRB} is similar to the observed one \cite{Hicks2}. However, the situation is more complex. A careful study of the magnetic structure using neutron scattering revealed alternating chirality in the chromium ion layers \cite{Takatsu2}, so the total chirality induced in the palladium layers by an adjacent pair of chromium layers is zero. On the other hand, nonreciprocity of electron transport along the layers in an external magnetic field was observed in \ch{PdCrO2} \cite{Akaike}, a necessary condition for which is a non-zero chirality of the effective field in the conducting layers. A possible resolution to this contradiction is provided by the measurements of short-range correlations, which demonstrated that the coherence length of the magnetic order along the $c$-axis (i.e., perpendicular to the layers) is only about 5.5 magnetic unit cells \cite{Billington}. This means that approximately 20\% of the unit cells contain layers with an effective field of non-zero chirality. It should be noted that \ch{PdCrO2} exhibits the highest conductivity among compounds with noncollinear magnetic order \cite{Mackenzie}. 

Another possibility for realizing the unusual half metals is offered by van der Waals layered heterostructures, in which an alternating sequence of magnetic (with helical ordering) and conductive hexagonal layers can be realized \cite{Park,Ju,Akatsuka,Soriano}. For example, the dichalcogenide \ch{NiI2} has a magnetic structure similar to that of the $p$-wave half-metal model discussed above \cite{Ju,Olsen}. A neutron diffraction study of bulk layered hexagonal \ch{VX2} (X = Cl, Br, and I) revealed the same 120$^\circ$ magnetic order in the base planes, as was discussed for $f$-wave half metals, with critical temperatures from 16~K to 36~K \cite{Hirakawa}. Later, it was shown that this magnetic structure can be stabilized in a \ch{VX2} monolayer even at much higher temperatures \cite{Wasey}. To obtain the unusual half metal, the Fermi level should be adjusted so that it lies within the half-metallic gap of the $p$- or $f$-wave type.   

\begin{figure*}
\centering	
 \includegraphics[scale=0.48]{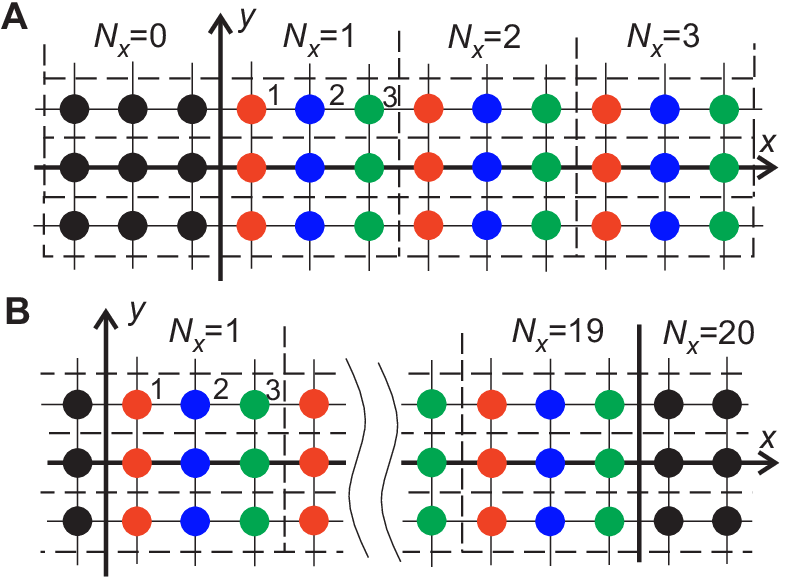}  \; \;  \; \;
 \includegraphics[scale=0.26]{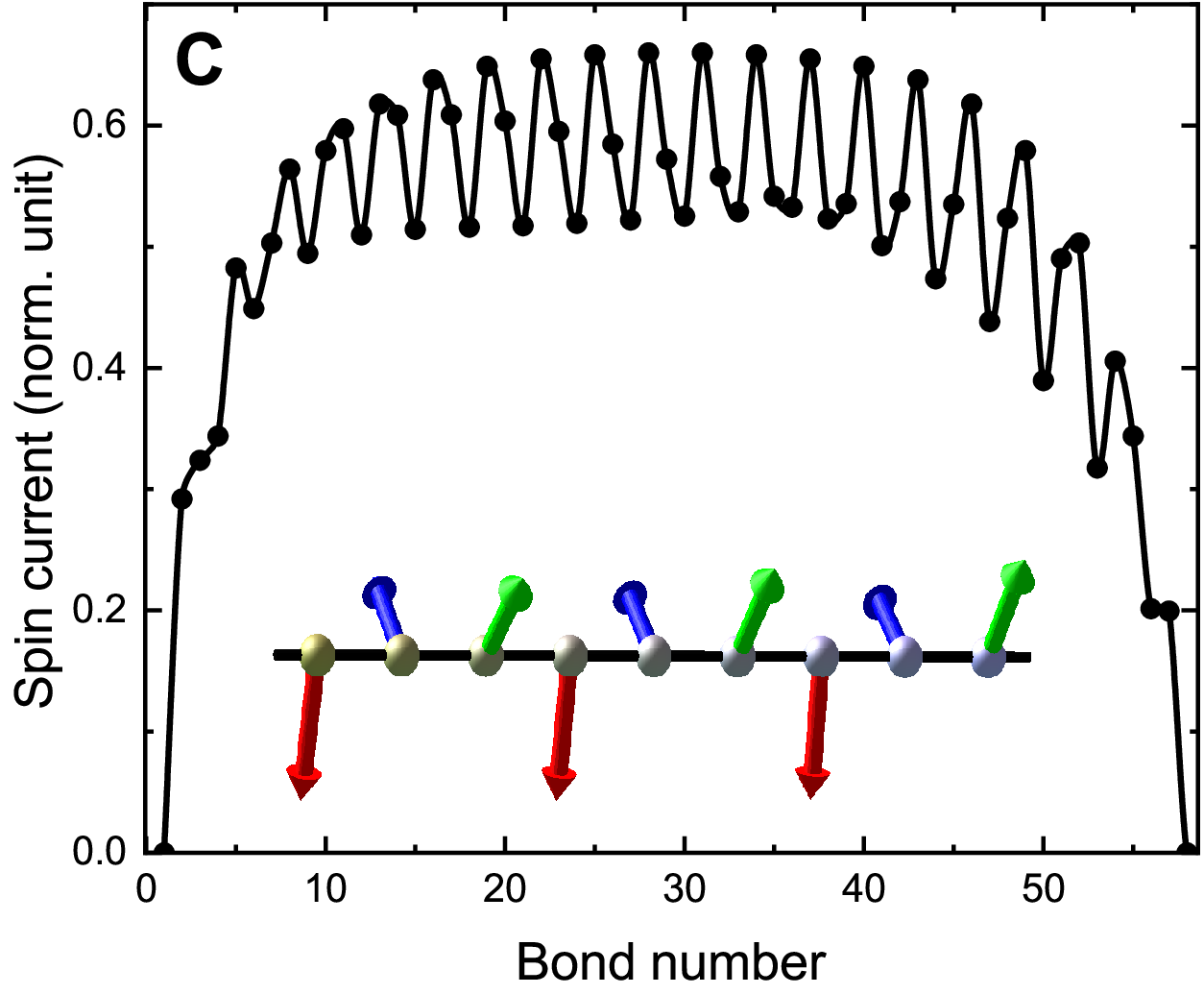} 
	\caption{\label{f4} (A) Schematic view of an infinite potential wall for the tight-binding rectangular model. The site colors denote the superlattices, the black color indicates the sites under the infinite potential. (B) Schematic view of the quantum well. (C) Persistent spin current in the quantum well. The helical spin structure in the central part of the quantum well is shown.}
\end{figure*}

The final problem we address in this Letter concerns the behavior of the unusual half metals in nanostructures. Let us consider an infinite potential wall for the $p$-wave half metal discussed above, assumed to be perpendicular to the $\mathbf{x}$ axis with $U(x<0) \rightarrow \infty$. Introducing a numbering of supercells along $\mathbf{x}$ by $N_x$, the negative values of $x$ correspond to $N_x \le 0$, as shown in Fig.~\ref{f4}A. Since the structure remains translationally invariant along the $\mathbf{y}$ axis, motion in this direction is the same as for a free particle. Then, the Hamiltonian (\ref{Htb}) for the $p$-wave model separates into two parts, and the problem reduces to one-dimensional motion along the $x$-axis.

Let the particle energy $E$ lie in the range marked by horizontal dashed lines in Fig.~\ref{f2}B. It should be noted that, for this given $E$, there are only two states instead of four (which exist for energies outside this range), namely, a left-moving particle with spin-up polarization and a right-moving particle with spin-down polarization. A superposition of these two states cannot yield zero amplitude at a lattice site and therefore cannot satisfy the boundary condition, because the basis for the wave function is incomplete. This is a well-known problem of complex band structure \cite{Chang,Dy}. It is resolved for the tight-binding model in section II of the SI, which results in two additional states with complex wave vectors. They take the form of evanescent waves with a helical spin structure. Thus, in the case of the infinitely high potential wall, the evanescent wave couples the incident and reflected waves with opposite spins.

Applying the full set of wave functions, we can investigate problems for a bounded $p$-wave half metal. For instance, in a quantum well (Fig.~\ref{f4}B) a standing wave with a helical spin structure is formed as shown in the inset of Fig.~\ref{f4}C. An operator for spin current can be derived, as usual, from the continuity equation (section III of the SI). It should be noted that the spinor potential, i.e., the second term in Hamiltonian Eq.~(\ref{Htb}), does not commute with the spin-projection operator. As in the Rashba medium \cite{Nikolic}, this results in a source of spin current in the equation and a persistent spin current in the quantum well, as shown in Fig.~\ref{f4}C. 

In summary, we demonstrate that a half-metallic gap can acquire a "time-reversal" nature---that is, be symmetric under simultaneous altering the wave vector direction and permutation of the spin-up and spin-down density of states. The $p$- and $f$-wave 2D half-metallic states can be realized. They have nodal points at the Fermi surface where the gaps for both spins either open, as in the $p$-wave state, or collapse, as in the $f$-wave one. The systems ($p$- and $f$-wave) considered in this Letter are altermagnets because they break the parity-time symmetry. To exhibit an unusual half-metallic state, the additional condition of the nontrivial topology is necessary. Since nonspin-flip scattering is strongly suppressed, anomalously high conductivity is anticipated.
A standing wave with a helical spin density and a persistent spin current arise in a quantum well based on these unusual half metals. Nanostructures based on the unusual $p$- and $f$-wave half metals can be applied in spintronics devices as spin valves \cite{Yuan} and others \cite{Ustinov}.

\section*{Acknowledgements}

This work was performed within the framework of the scientific program of the National Center for Physics and Mathematics (Section No. 7 “Investigations in high and ultrahigh magnetic fields", 2026-2028).

\section{Supporting information}
The Supporting Information is available free of charge at [URL].
Explicit forms of the 2D tight-binding Hamiltonians for $p$- and $f$-wave systems (section I), a solution for the complex band structure in the $p$-wave half-metallic state (section II), a solution for $p$-wave half metal in a quantum well (section III), and the definition of spin current for the tight-binding model (section IV) (PDF).


\bibliography{unusual_b}

\end{document}